\documentclass[aps,prb,amsmath,amssymb,twocolumn,superscriptaddress]{revtex4-1}

\usepackage{graphicx}
\usepackage{dcolumn}
\usepackage{bm}

\setcitestyle{numbers,square}

\begin{document}

\title{Effect of short-range correlations on spectral properties of doped Mott insulators}

\author{V. I. Kuz'min}
\email{kuz@iph.krasn.ru}
\affiliation{Kirensky Institute of Physics, Federal Research Center KSC SB RAS, Krasnoyarsk, 660036 Russia}

\author{S. V. Nikolaev}
\affiliation{Kirensky Institute of Physics, Federal Research Center KSC SB RAS, Krasnoyarsk, 660036 Russia}
\affiliation{Siberian Federal University, Krasnoyarsk, 660041 Russia}

\author{S. G. Ovchinnikov}
\affiliation{Kirensky Institute of Physics, Federal Research Center KSC SB RAS, Krasnoyarsk, 660036 Russia}
\affiliation{Siberian Federal University, Krasnoyarsk, 660041 Russia}

\date{\today}

\begin{abstract}
In the framework of cluster perturbation theory for the 2D Hubbard and Hubbard-Holstein models at low hole doping we have studied the effect of local and short-range correlations in strongly correlated systems on the anomalous features in the electronic spectrum by investigating the fine structure of quasiparticle bands. Different anomalous features of spectrum are obtained as the result of intrinsic properties of strongly correlated electron and polaron bands in the presence of short-range correlations. Particularly, features similar to the electron-like Fermi-pockets of cuprates at hole doping $p\sim0.1$ are obtained without \textit{ad hoc} introducing a charge density wave order parameter within the Hubbard model in a unified manner with other known peculiarities of the pseudogap phase like Fermi-arcs, pockets, waterfalls, and kink-like features. The Fermi surface is mainly formed by dispersive quasiparticle bands with large spectral weight, formed by coherent low-energy exications. Within the Hubbard-Holstein model at moderate phonon frequencies we show that modest values of local electron-phonon interaction are capable of introducing low-energy kink-like features and affecting the Fermi surface by hybridization of the fermionic quasiparticle bands with the Franck-Condon resonances.
\end{abstract}

\maketitle

\section{\label{Intro}Introduction} 
The physics of strongly correlated  electron systems is greatly influenced by many-particle correlation effects. They result in complicated phase diagrams and exotic properties of strongly correlated compounds. A significant contribution to the formation of such effects is provided by local and short-range correlations as the consequence of Coulomb interaction's local character. One of the manifestations of correlation effects is the presence of different anomalies in the electronic spectrum. Due to the ability to investigate the electronic structure by means of angle-resolved photoemission spectroscopy (ARPES), anomalous properties of high-${T}_{c}$ cuprates, such as the pseudogap and Fermi-arcs, are known \cite{Damascelli03}. Further ARPES studies of high-${T}_{c}$ cuprates, as well as other strongly correlated compounds, revealed a variety of low-energy \cite{Campuzano99, Lanzara01, Lanzara10, Kaminski01, Johnson01, Sato03, Gweon06, Park08, He13, Plumb13, Anzai17} and also high-energy \cite{Graf07, Meevasana07, Zhang08, Ikeda09, Moritz09} kinks along with the so-called waterfalls. Another anomalous property, an electron-like pocket at the Fermi surface at hole doping $p\sim0.1$, has been uncovered by the observation of quantum oscillations \cite{DL07, Barisic13} in conjunction with the measurements of the Hall, Seebeck, and Nernst coefficients \cite{LeBoeuf07, DL13} in high-${T}_{c}$ hole-doped cuprates. A change in the Hall coefficient's sign has been detected in ${\text{YBa}_{2}\text{Cu}_{3}\text{O}_{y}}$ at $p\approx0.08$ \cite{LeBoeuf11}, implying a Fermi surface reconstruction. Hole-like pockets have also been reported \cite{DL15} to coexist with an electron pocket in ${\text{YBa}_{2}\text{Cu}_{3}\text{O}_{y}}$.

At present there is no consensus on the nature of all these anomalies in the electronic structure of strongly correlated systems. However, short-range correlations stemming from Coulomb repulsion in correlated metals in the vicinity of the Mott transition are known to produce significant influence on the pseudogap and Fermi-arcs \cite{Senechal04, Stanescu06, Korshunov07}. Moving away from the Fermi energy we face low-energy kinks, which are usually observed at energies $\lesssim100\text{meV}$, and high-energy kinks, at energies $\gtrsim500\text{meV}$, accompanied by waterfalls. There is a number of works in which low-energy kink-like features have been modeled in terms of perturbation theory \cite{Eschrig00, Eschrig03, Chubukov04, Devereaux04, Sandvik04, Dahm09, Mazur10, Mazza13, Geffroy16} due to the interaction of electrons with phonons and (or) spin fluctuations. In Ref.~\onlinecite{Mazur10} high-energy kinks due to phonons have also been reported. However, it appears that the presence of some special electron-boson interaction is not necessary to observe kink-like features in the electronic structure, as it was shown within dynamical mean-field theory (DMFT) for the Hubbard model \cite{Byczuk07}. This way, in the framework of $\text{DMFT}+\Sigma$ \cite{Sadovskii11} and within DMFT simulations of the Hubbard-Holstein model \cite{Bauer10} both phononic and pure electronic low-energy kinks have been observed. Low-energy anomalies have been considered within diagrammatic quantum Monte Carlo for the t-J-Holstein model \cite{Mishchenko11}. Strong evidence in support of the important role of local Coulomb interaction in the formation of high-energy waterfall anomalies has been obtained by dynamical cluster \cite{Macridin07} and determinant quantum Monte-Carlo (DQMC)\cite{Moritz09}. The whole picture of the electronic structure's features, such as Fermi-arcs, kinks, and waterfalls, has been obtained within the cluster DMFT \cite{Sakai10} and cluster perturbation theory (CPT) \cite{Kohno12, Kohno14} studies of the Hubbard model, as well as within CPT applied to the t-J model \cite{Kohno15}. The influence of correlations and spin fluctuations on the features of single-electron spectrum have been studied within the framework of CPT applied to the Hubbard model, the t-J model, and the t-J model with three-site correlated hoppings \cite{Wang15, Wang_arxiv}. The comparison of the electronic spectral function of the Hubbard and t-J models reveals important influence of three-site correlated hoppings on the high-energy electronic structure \cite{Wang15, Kuzmin14}. Aside from one-band models, features of electronic spectrum have been recently studied within the three-band model of cuprates using DQMC, exact diagonalization (ED) and CPT \cite{Wang16}, providing information about contribution from different orbitals to the spectral function. Turning back to the Fermi level, the Fermi surface's reconstruction due to the charge density wave (CDW) ordering at $p\sim0.1$ was suggested as the mechanism of the nodal electron pocket's formation \cite{Harrison11} and shown to produce coexisting electron and hole pockets within the phenomenological mean-field model \cite{Allais14} in agreement with the finding in the experiments on quantum oscillations \cite{DL15}. However, within a single-electron approach, the presence of density waves is the only mechanism that can lead to the Fermi-surface reconstruction of this type. As opposed to this, the quasiparticle bands in the presence of short-range strong correlations behave in a highly nontrivial way.

The aim of this paper is to study the role played by local and short-range correlations stemming from electron-electron and electron-phonon interactions on the formation of anomalous features in the electronic structure of strongly correlated electron systems within the 2D Hubbard and Hubbard-Holstein models at hole doping. The Hubbard model is the fundamental microscopic model to investigate the role of correlations due to on-site Coulomb repulsion. It can be obtained as the effective model of cuprates for the excitation energy $E < E_t$, where $E_t\approx 2eV$ is the two-hole triplet state energy over the Zhang-Rice singlet \cite{Korshunov05}. The Hubbard-Holstein model includes the interaction of phonons with local electron density over the Hubbard Hamiltonian. 

We aim to obtain the fine structure of quasiparticle bands contributing to the electron spectral function within CPT \cite{Senechal00, Senechal02}. CPT provides a possibility to account exactly for  local and short-range correlations in the framework of a finite cluster. Usually CPT is based on the exact cluster ground state and a few excited states obtained by the Lanczos method \cite{Dagotto94}. In a strongly correlated system even high-energy states can be important to obtain the excitation spectrum. Thus, we use full ED instead of the Lanczos method, to calculate cluster's Green's functions. The side benefit of full ED approach is the ability to calculate the electronic structure without an artificial Lorentzian broadening of the spectral function within a norm-conserving approximation. However, implementations of CPT based on Lanczos method are capable of treating square clusters of 16 sites in the Hubbard model, while the largest square cluster accessible by full ED is 9-site one. Nevertheless, we will show that the spectral function obtained within CPT implemented with 9-site with some Lorentzian broadening is in a good agreement with the results of CPT on a 16-site cluster, obtained with the same broadening. Thus, we have a starting point to study the fine structure of the spectrum. Working with the Hubbard-Holstein model within the same approach we use a 4-site square cluster with 8 phonons. With respect to the Hubbard-Holstein model the dignity of CPT is its ability to treat contributions from electron-electron and electron-phonon interactions to the short-range and local correlations on an equal footing.

This paper is organized as follows. In section \ref{sec:Models} we provide information about the models under consideration as well as a brief discussion of CPT for the convenience of the reader. Sections \ref{sec:H} and \ref{sec:HH} present our results on the anomalous spectral features of the Hubbard and Hubbard-Holstein models. In section \ref{sec:Conclusion} we give concluding remarks.

\section{\label{sec:Models}Models and method} 

The Hubbard model \cite{Hubbard63} is given by the Hamiltonian  
\begin {equation}
H=\sum\limits_{i,\sigma }{\left\{{\left({\varepsilon-\mu} \right){n_{i,\sigma }}+\frac{U}{2}{n_{i,\sigma }}{n_{i,\bar\sigma}}} \right\}-\sum\limits_{ij,\sigma} t_{ij} a_{i,\sigma }^{\dag} a_{j,\sigma}^{}},
\label {eq:1}
\end {equation}
where $a_{i,\sigma }$ is the annihilation operator of an electron on a site $i$, $n_{i\sigma}=a_{i\sigma}^{\dag}a_{i\sigma}^{}$, $t_{ij}$ is the hopping integral, $U$ is the on-site Coulomb interaction. The Hubbard-Holstein model is obtained from Eq.~\ref{eq:1} by adding the optical phonon energy and local electron-phonon interaction terms \cite{Holstein59_1, Holstein59_2}
\begin {equation}
H'= \omega_{ph}\sum\limits_{i}b_{i}^{\dag}b_{i}^{} - g\sum\limits_{i}{n_{i}}\left(b_{i}^{\dag} + b_{i}^{}\right).
\label {eq:2}
\end {equation}
In Eq.~\ref{eq:2} the phonon annihilation operators $b_{i}$ are introduced, $\omega_{ph}$ is the phonon frequency in the units of $\hbar$, $g$ is the electron-phonon coupling constant, and $n_i = n_{i\uparrow} + n_{i\downarrow}$. We introduce the dimensionless coupling constant $\lambda = 2g^2/\left(\omega_{ph}W\right)$, where $W=8t$ is the noninteracting bandwidth.

The first step in CPT construction is to cover the lattice by translations of a cluster. The main idea of CPT is to account for the short-range correlations explicitly while considering long-range interactions in terms of perturbation theory \cite{Ovchinnikov89, Gros93}. In its present form it was derived from strong-coupling perturbation theory \cite{Senechal02}. It was shown \cite{Potthoff03} to be a limiting case of the self-energy-functional approach \cite{Potthoff03_e}. It is also convenient to implement CPT by working within the formalism of the Hubbard X-operators constructed from exact cluster eigenstates for two reasons \cite{Nikolaev10}. First, diagonal X-operators give the local eigenstates, while off-diagonal describe Fermi- and Bose-type excitations. That allows to rewrite the intercluster interactions as a bilinear product of X-operators. Second, in spite of large number of local eigenstates, $4^{N_c}$ for a cluster with $N_c$ atoms, in the case of the Hubbard model, only the small part of $4^{2N_c}$ local quasiparticles give significant contribution to the electron spectral function. This allows to formulate a norm conserving approximation. To control the sum rule for the spectral weight function $A_{\sigma}\left(\mathbf{k}, \omega \right)$ we denote $f = \int{d\omega}A_{\sigma}\left(\mathbf{k}, \omega \right)$. Taking all excitations into account one obtains $f=1$. For example, for a 2x2 cluster at half filling and $U=8t$ with account for only near-neighbor hoppings, the modest number of 25 excitations provides $f>0.9999$. Working with 3x3 cluster at low doping levels, it is usually required to account for less than 2000 excitations to obtain $f>0.997$. We introduce a two-dimensional index $\alpha = \left(p, q\right)$ that numerates excitations with annihilation of an electron. Then we built Hubbard operators $X^{\alpha} = {\left|p\right\rangle}{\left\langle q \right|}$ \cite{Hubbard65} on the basis of cluster eigenstates obtained by full ED. X-operators provide a natural formalism to represent an electron as a composite quasiparticle: the annihilation operator of an electron with spin $\sigma$ on a site $i$ is a linear combination of X-operators, $c_{\sigma i}=\sum_{\alpha}{\gamma_{\sigma i}\left(\alpha\right)X^{\alpha}}$, where $\gamma_{i}\left(\alpha\right)$ are the annihilation operator's matrix elements. Both Hamiltonians under consideration can be rewritten as the sum of the intracluster part and intercluster hopping: 
\begin {equation}
H= \sum\limits_{\mathbf{f},n}{E_\mathbf{n} X^{nn}} + \sum\limits_{\mathbf{f},\mathbf{\Delta}>0}{T_{\mathbf{\Delta}}^{\alpha,\beta}{X_{\mathbf{f}}^{\alpha}}^{\dag}X_{\mathbf{f} + \mathbf{\Delta}}^{\beta}},
\label {eq:3}
\end {equation}
where $\mathbf{f}$ is a cluster coordinate, $E_n$ is the exact cluster eigenstate, $\mathbf{\Delta}$ is a vector connecting nearest clusters, $T_{\mathbf{\Delta}}^{\alpha,\beta}$ describes intercluster hopping of excitations.

Next, we define the Green's functions $D_{\alpha,\beta}\left(\mathbf{\tilde{k}},w\right) = {\left\langle \left\langle X^{\alpha}|{X^{\beta}}^{\dag} \right\rangle \right\rangle}_{\mathbf{\tilde{k}},\omega}$, where $\mathbf{\tilde{k}}$ is the wave vector defined in the reduced Brillouin zone. The generalized Dyson equation derived in terms of the Hubbard operators diagram technique \cite{Zaitsev75, Izyumov91} reads \cite{book_Valkov}:
\begin {eqnarray}
\hat{D}\left(\mathbf{\tilde{k}},\omega\right) 
= [{\hat{G^0}\left(\omega\right)}^{-1}
&-&\hat{P}\left(\mathbf{\tilde{k}},\omega\right)\hat{T}\left(\mathbf{\tilde{k}}\right) \nonumber \\
&+& \hat{\Sigma}\left(\mathbf{\tilde{k}},\omega\right)
 ]^{-1}\hat{P}\left(\mathbf{\tilde{k}},\omega\right),
\label {eq:4}
\end {eqnarray} 
where
\begin {equation}
T_{\alpha \beta}\left(\mathbf{\tilde{k}}\right) = \sum\limits_{\mathbf{\Delta>0}}
\left(T_{\mathbf{\Delta}}^{\alpha\beta}e^{i\mathbf{\tilde{k}}\mathbf{\Delta}} 
- T_{\mathbf{\Delta}}^{\beta\alpha}e^{-i\mathbf{\tilde{k}}\mathbf{\Delta}}\right)
\label {eq:5}
\end {equation}
is the element of the hopping matrix and  
\begin {equation}
{G^0_{\alpha,\beta}\left(\omega\right)} = \frac{\delta_{\alpha,\beta}}{\omega - E_{\alpha} + \mu}
\label {eq:6}
\end {equation}
is the exact local propagator, $E_{\alpha} = E_{q} - E_{p}$, and $\mu$ is the chemical potential. In Eq.~\ref{eq:4}, $\hat{\Sigma}\left(\mathbf{q},\omega\right)$ is the intercluster self-energy and $\hat{P}\left(\mathbf{\tilde{k}},\omega\right)$ is the strength operator. In the Hubbard-I approximation for the intercluster hopping one has $\hat{\Sigma}\left(\mathbf{\tilde{k}},\omega\right) = 0$ and $P_{\alpha \beta}\left(\mathbf{\tilde{k}},\omega\right) = \delta_{\alpha \beta} F\left(\alpha\right) = \delta_{\alpha \beta} \left(\left\langle X^{pp} \right\rangle + \left\langle X^{qq} \right\rangle \right)$. In this approximation Eq.~\ref{eq:4} reduces to the CPT matrix equation \cite{Senechal00}, although in band representation,
\begin {equation}
\hat{D}\left(\mathbf{\tilde{k}},\omega\right)^{-1} = \hat{D^0}\left(\omega\right)^{-1} - \hat{T}\left(\mathbf{\tilde{k}} \right),
\label {eq:7}
\end {equation}
where ${D^0}_{\alpha\beta}\left(\omega\right) = F_{\alpha}{G^0}_{\alpha\beta}\left(\omega\right)$.  

Finally, the translation-invariant electron Green's function is recovered following Ref.~\onlinecite{Senechal02}:
\begin {eqnarray}
&G_{\sigma}\left(\mathbf{k}, \omega\right) = \frac{1}{N_c} \nonumber \\
\times& \sum\limits_{\alpha\beta} {\sum\limits_{ij} {\gamma_{\sigma i}\left(\alpha\right)
\gamma_{\sigma j}\left(\beta\right) e^{-i\mathbf{k}\left(\mathbf{r}_i - \mathbf{r}_j \right)}
D_{\alpha\beta}\left(\mathbf{k},\omega\right)}},
\label {eq:8}
\end {eqnarray} 
where $N_c$ is the number of sites within a cluster and $\mathbf{k}$ is defined in the original Brillouin zone.

Thus, the spectral function $A_{\sigma}\left(\mathbf{k}, \omega\right) = - \frac{1}{\pi}\text{Re}G_{\sigma}\left(\mathbf{k}, \omega\right)$ is distributed among the bands of so-called Hubbard fermions (or polarons), defined by the poles of Eq.~\ref{eq:8}, and for a given wave number may be approximated by the Lorentzian distribution. In the Landau Fermi-liquid theory a quasiparticle is coherent when its damping is small and spectral intensity is close to unity. Although the imaginary part of the self-energy is zero in CPT approximation, some qualitative information about the degree of quasiparticle coherence/incoherence can be gained from the spectral intensity. Single dispersive quasiparticle bands with large spectral weight (LSW) $A_{\sigma}\left(\mathbf{k}, \omega\right) \sim 0.5 $ can be interpreted as coherent parts of the spectrum, while incoherent excitations are represented by multiplicity of weakly dispersive bands with small spectral weight (SSW) $A_{\sigma}\left(\mathbf{k}, \omega\right) \ll 0.5 $ as the result of a decay of an electron. A Lorentzian broadening of spectral delta peaks is often used in CPT. It allows to make a transformation from a descrete set of quasiparticle bands to a continuous energy distribution of  $A_{\sigma}\left(\mathbf{k}, \omega\right)$, like in an infinite system, in order not to overemphasize the salient features of regions with multiple SSW bands. Also, it allows to qualitatively model the broadening of ARPES spectra, which depends on different factors like finite resolution, averaging over energy window, or temperature.  However, in the regions consisting of single LSW quasiparticles important fine features can be lost at the same time. So, we find it useful to discuss the bare (without a Lorentzian broadening) CPT result, as well as the representation in which a finite Lorentzian broadening is used.

\section{\label{sec:H}the Hubbard model}

We now discuss our results on the spectral function of the 2D Hubbard $t-t'-t''-U$ model, where $t$, $t'$, and $t''$ stand for the hopping integrals between the first, the second, and the third neighbors. The energy $\omega$ is measured in the units of $t$. The $U=8t$ value of Coulomb repulsion is fixed. To plot the results without a Lorentzian broadening, a stepwise broadening of halfwidth $0.04t$ is given to the spectral lines to make them clearly visible.

\begin{figure}
\includegraphics{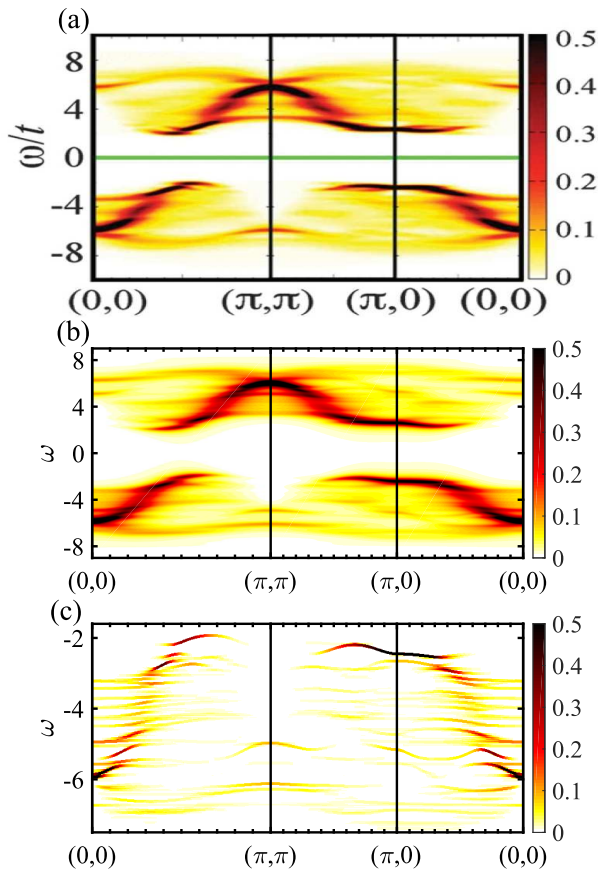}
\caption{\label{fig:Fig1} The spectral weight distribution obtained within the CPT for the 2D Hubbard model at $p=0$, $t'=0$, $t''=0$ using (a) 4x4 cluster (adapted from Ref.~\onlinecite{Kohno12}), (b),(c) 3x3 cluster. In (b) the Lorentzian broadening is $\delta = 0.16t$, in (c) no Lorentzian broadening is used. In (c) only LHB is shown.}
\end{figure}

Figure~\ref{fig:Fig1}(b) displays our results for the spectral function plotted with a Lorentzian broadening $\delta = 0. 16t$ at hole doping $p=0$ and only nearest hoppings taken into account in comparison with the results calculated for the same parameters within 4x4 cluster CPT in Ref.~\onlinecite{Kohno12} (see Fig~\ref{fig:Fig1}(a)). General agreement between (a) and (b) in Fig.~\ref{fig:Fig1} is seen. By comparing panels (b) and (c) of Fig.~\ref{fig:Fig1}, one can see the correspondence between the broadened spectral weights and the representation of the bare CPT result in Fig.~\ref{fig:Fig1}(c), where $\sim10^3$ quasiparticle bands have been accounted for (the spectral sum $f$ is respected as $f>0.997$ here and below). For concreteness we concentrate now on the low Hubbard band (LHB). Particularly, the most important for the following discussion is that in the $\left(0,0\right)-\left(\pi,\pi\right)$ direction the low energy mode at $\omega\approx-2t$ is present, and it is isolated from the more high-energy mode at $-3.5t\lesssim\omega\lesssim-2t$ by the kink-like feature (see Fig.~\ref{fig:Fig1}(a, b)). Around the $\left(\pi,0\right)$ point there is the flat mode. In Fig.~\ref{fig:Fig1}(c) it is seen that both the low-energy mode and the flat mode are single LSW quasiparticles bands and  the low-energy kink-like feature is due to the energy gap, which is seen right below the low-energy mode in the bare CPT result. This kink-like feature can be viewed as a point dividing the energy regions with different properties of quasiparticle bands: the more high energy modes at $-3.5t\lesssim\omega\lesssim-2t$ both in $\left(0,0\right)-\left(\pi,\pi\right)$ and $\left(\pi,0\right)-\left(0,0\right)$ directions consist of several bands (see Fig.~\ref{fig:Fig1}(c)).
Both in $\left(0,0\right)-\left(\pi,\pi\right)$ and $\left(\pi,0\right)-\left(0,0\right)$ directions the high-energy kink behavior at $\omega\sim-3.5t$ with waterfall-like features below is observed in Fig.~\ref{fig:Fig1}(b). Compared to the results with a 4x4 cluster, the waterfall-like anomalies in Fig.~\ref{fig:Fig1}(b) have artificial kink-like features inside due to the lack of relevant quasiparticle bands obtained with a 3x3 cluster. From considering Fig.~\ref{fig:Fig1}(c) it is seen that the waterfall-like region $-5t\lesssim\omega\lesssim-3.5t$ of Fig.~\ref{fig:Fig1}(b) is formed by a large number of weakly dispersive SSW bands. So, in Fig.~\ref{fig:Fig1}(b), the feature at $\omega\sim-3.5t$, which is reminiscent of a high-energy kink, can be viewed as dividing the modes consisting of several bands at $\omega \sim -3t$ from the multiplicity of SSW bands. Notably, a LSW band is present at $\omega\sim-6t$ in Fig.~\ref{fig:Fig1}(c) around $(0,0)$, resulting in a LSW in the corresponding region (first noticed in Ref. \onlinecite{Preuss95}) in Fig.~\ref{fig:Fig1}(b) . Also, at $\omega\lesssim-6t$ the satellite mode is present in Fig.~\ref{fig:Fig1}(b), formed by a number of weak dispersive bands (see Fig.~\ref{fig:Fig1}(c)).

\begin{figure}
\includegraphics{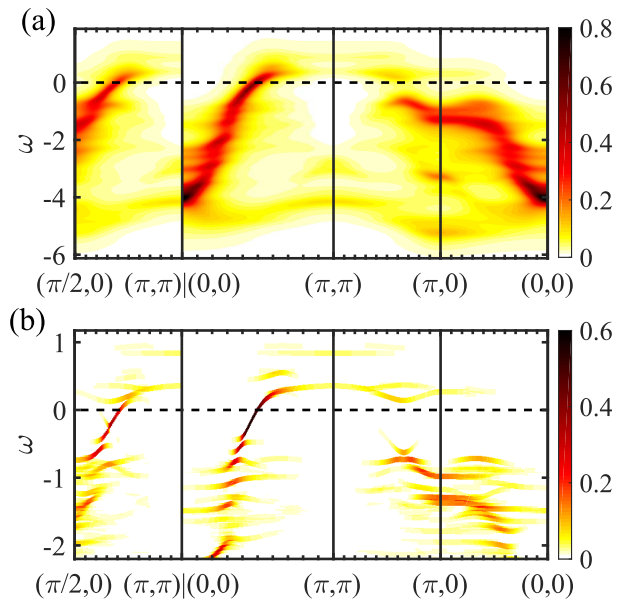}
\caption{\label{fig:Fig2} (a),(b) The spectral weight distribution in the LHB at $p=0.03$, $t'=-0.2t$, $t''=0.15t$ (a) with $\delta = 0.16t$, (b) without a Lorentzian broadening. The dashed line indicates the Fermi level.}
\end{figure} 

Based on the comparison above, we conclude that the main features of our results within a 3x3 cluster CPT are in a good agreement with 4x4 CPT, and they can be useful when an access to fine properties of quasiparticle bands is needed. From general remarks concerning the electronic structure of the 2D Hubbard model we now turn to the case more relevant to the pseudogap phase of cuprates by accounting for non-zero hole doping and typical hopping integrals $t'$, $t''$, which are similar to the estimates made for $\text{La}_{2-x}\text{Sr}_x\text{CuO}_4$ at low doping \cite{Yoshida06}. We fix $t'=-0.2t$, $t''=0.15t$, for the following discussion, but the main results are stable with respect to moderate variations of the parameters.

\begin{figure}
\includegraphics{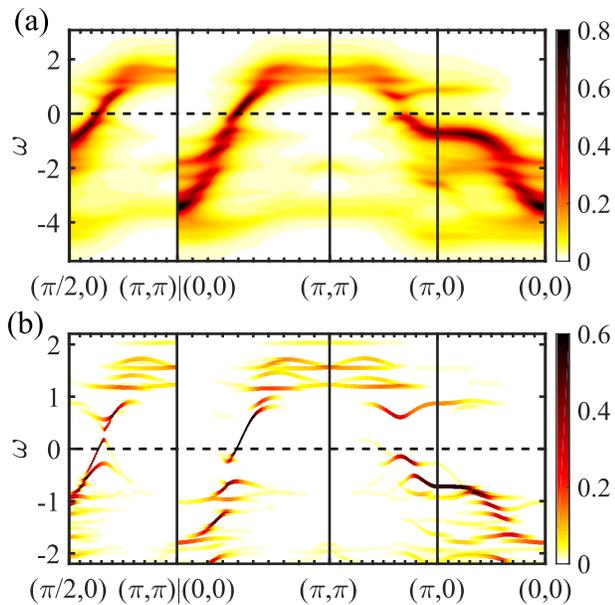}
\caption{\label{fig:Fig3} The same as in Fig.~\ref{fig:Fig2} at $p=1/9$.}
\end{figure}

\begin{figure}
\includegraphics{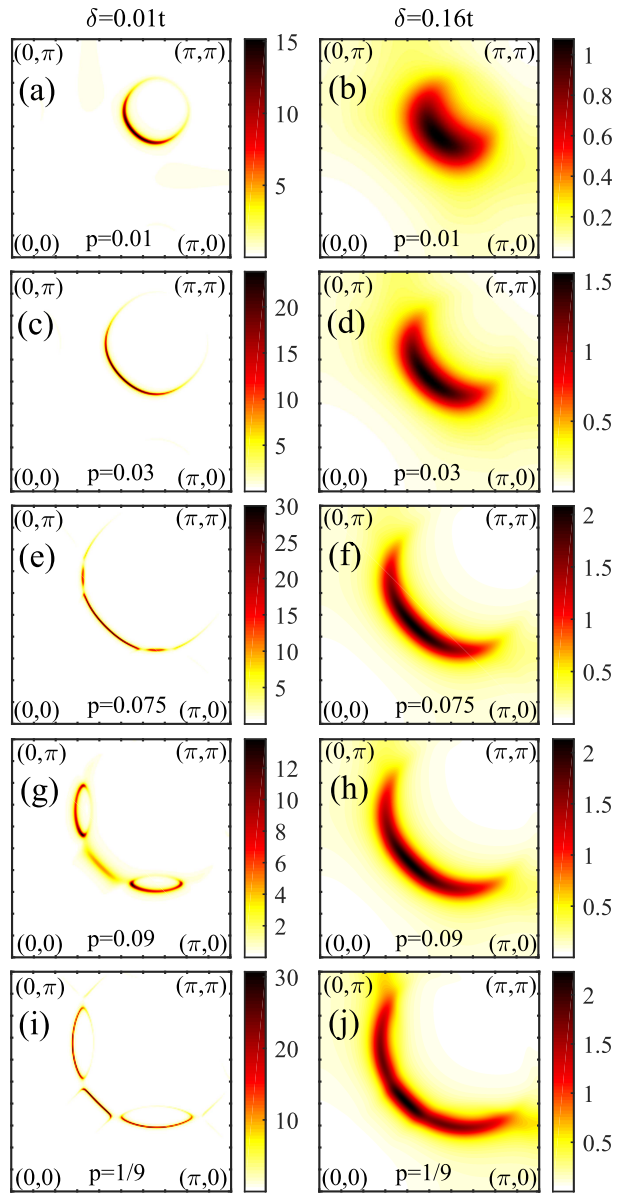}
\caption{\label{fig:Fig4} $A_{\sigma}\left(\mathbf{k}, \omega = 0\right)$ at various doping levels $p$ plotted using two different values of a Lorentzian broadening $\delta$.}
\end{figure}

From Fig.~\ref{fig:Fig2}, which shows the spectral function at low doping $p=0.03$ with account for non-nearest hopping, it is seen that the flat mode is moved towards $\omega \sim -t$ and becomes incoherent , while the coherent low-energy band survives, as shown by presenting the momentum cuts in $\left(0,0\right)-\left(\pi,\pi\right)$ and $\left(\pi/2,0\right)-\left(\pi,\pi\right)$ directions. Thus, while the plot of the spectral function with a significant broadening shows a behavior of a Fermi-arc, which grows with doping (see Fig.~\ref{fig:Fig4}(b),(d)), in fact, the Fermi surface at values of doping $p\lesssim0.075$ is a pocket with a non-uniform distribution of the spectral weight along the Fermi contour (see Fig.~\ref{fig:Fig4}(a),(c)).

An interesting effect is seen with further doping: while the Fermi level moves through the low-energy band towards a nodal kink-like feature in Fig.~\ref{fig:Fig3}(a), the hole-like Fermi pocket in the nodal direction undergoes a transition towards an electron pocket, as it is seen from the dispersion in Fig.~\ref{fig:Fig3}(b). This way, the Fermi surface undergoes a topological transition at $p\approx0.075$ (see Fig.~\ref{fig:Fig4}(e)). The local electron-like character of dispersion is not clearly seen in Fig.~\ref{fig:Fig3}(a), where the spectral peaks are smeared out. In $\left(\pi/2,0\right)-\left(\pi,\pi\right)$ direction a well-defined quasiparticle band emerges at the Fermi energy. As a result, a nodal electron pocket at the Fermi surface coexists with two hole-like pockets, elongated along $\left(0,0\right)-\left(\pi,0\right)$ and $\left(0,0\right)-\left(0,\pi\right)$ directions, as it is shown in Fig.~\ref{fig:Fig4}(i), similar to the results within the phenomenological model of CDW ordering \cite{Allais14}. We note that Fermi-surfaces of this type have been obtained recently using CPT with 3x3 and 4x4 clusters within a t-J-type model (see Ref.~\onlinecite{Ivantsov17}). The Fermi surface evolution, when plotted with a large value of a Lorentzian broadening (see the right column of Fig.~\ref{fig:Fig4}), appears as the growth of an arc, which possibly explains why electron and hole pockets are not commonly detected by ARPES. 

\begin{figure}
\includegraphics{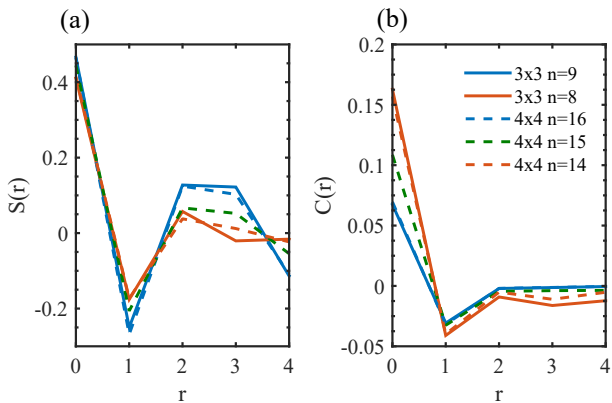}
\caption{\label{fig:Fig5} (a) Spin and (b) charge correlation functions calculated within clusters with the specified size and number of electrons $n$.}
\end{figure}

In our calculations, the existence of the nodal electron pocket is due to the low-energy band consisting of a single Hubbard fermion. By the construction of CPT, it results from the effect of the short-range correlations on the quasiparticle dispersion, without \textit{ad hoc} introduced CDW. Thus, the CDW argumentation may be an artifact resulting from a single-electron band structure approach \footnote{A. Sherman, a remark during discussion of the CDW mechanism of electronic pocket in hole doped cuprates at the Superstripes-2016 conference  in Ichia, Italy}. To further clarify the influence of short-range correlations in our calculations we study intracluster spin correlation functions
\begin {equation}
S\left(r\right) = \left\langle S_0^+S_r^- \right\rangle,
\label {eq:9}
\end {equation}
where $S_0^+=a_{0,\uparrow }^{\dag} a_{0,\downarrow}^{}$, $S_r^-=a_{r,\downarrow }^{\dag} a_{r,\uparrow}^{}$,  
and charge correlation functions
\begin {equation}
C\left(r\right) = \left\langle \left( n_0 - \left\langle n_0 \right\rangle \right)\left( n_r - \left\langle n_r \right\rangle \right) \right\rangle.
\label {eq:10}
\end {equation}
A site ``$0$'' in Eqs.~\ref{eq:9},~\ref{eq:10} is a corner site, and $r$ denotes a site belonging to the $r$-th coordinate sphere. Fig.~\ref{fig:Fig5} shows the correlation functions calculated within a 3x3 cluster in comparison with the results on 4x4 cluster obtained with the ground-state Lanczos method. A spin-liquid state, with $\left\langle \left(n_{i\uparrow}-n_{i\downarrow}\right)\right\rangle=0$ and short-range antiferromagnetic correlations decreasing with doping, is realized within a cluster. At the same time, the charge correlations do not show a tendency to form a density wave, only the contributions from the zeroth and the first coordinate spheres are significant. The values of correlation functions calculated within a 3x3 cluster are in a good agreement with the trends in the results on a 4x4 cluster.

\section{\label{sec:HH}the Hubbard-Holstein model}

\begin{figure}
\includegraphics{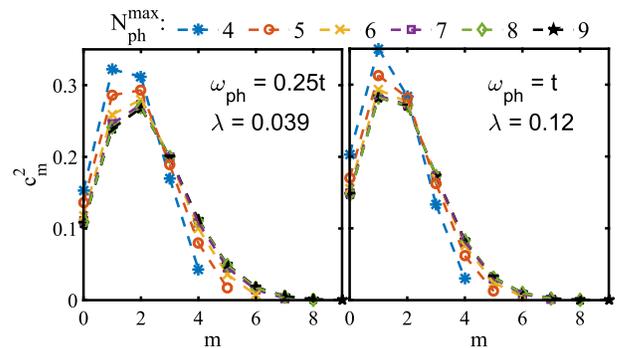}
\caption{\label{fig:Fig6} The distribution of $c_m^2$, defined in Eq.~\ref{eq:9}, obtained with fixed values of $N_{ph}^{max}$ from 4 to 9.}
\end{figure}

In this section we present our results on the Hubbard-Holstein model. As in the preceding section, we fix $t'=-0.2t$, $t''=0.15t$, $U=8t$. In our CPT calculations we were restricted with a 2x2 cluster and 8 phonons. Consider the ground state cluster wave function written as:
\begin{equation}
\left|\psi_0\right\rangle = \sum\limits_{f{p_m}}{c_{fp_m}\left|{e_f}\right\rangle\left|{e_{p_m}}\right\rangle}
= \sum\limits_{m}{c_m\left|{\phi_m}\right\rangle},
\label{eq:11}
\end{equation}
where $\left|{e_f}\right\rangle$ is a fermion basis state,  $\left|{e_{p_m}}\right\rangle$ are phonon basis states with $m$ phonons, and $\left|{\phi_m}\right\rangle$ denotes the $m$-phonon part of the wave function. Without electron-phonon interaction the only non-zero contribution is $c_0=1$. The maximum of the distribution of coefficients $c_m$, defined in Eq.~\ref{eq:11} generally shifts towards higher numbers of phonons while increasing the electron-phonon constant or lowering the frequency, this shift demonstrates the polaronic effect in Fig.~\ref{fig:Fig6}. Performing ED at fixed $\omega_{ph}$ we chose the maximal parameter $\lambda$ at which the change in the cluster's ground state energy does not exceed $10^{-3}$ while increasing the number of phonons $N_{ph}^{max}$ within the cluster from 8 to 9. In Fig.~\ref{fig:Fig6} we show the convergence of the distribution of $c_m^2$ with increasing $N_{ph}^{max}$ for two sets of parameters that will be used in the following discussion, $\omega_{ph} = 0.25t, \lambda = 0.039$ and $\omega_{ph} = 1t, \lambda = 0.12$.

\begin{figure}
\includegraphics{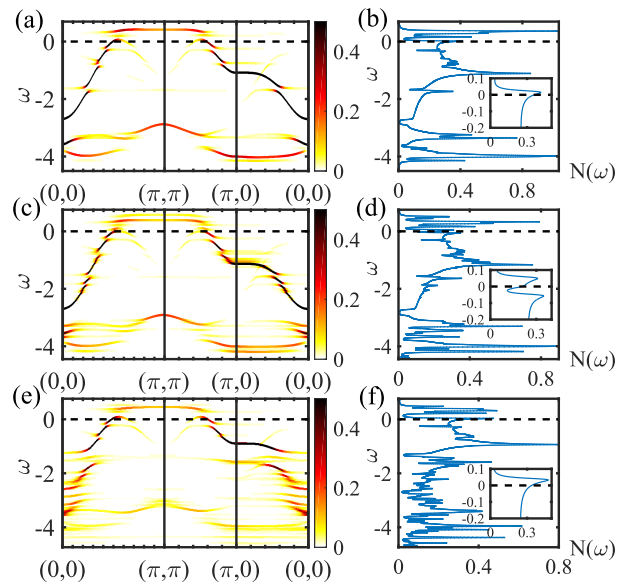}
\caption{\label{fig:Fig7} (a),(c),(e) The spectral weight distribution in the LHB plotted at $p=0.05$ doping without a Lorentzian broadening and (b),(d),(f) the corresponding density of states plotted with $\delta = 0.01t$ for (a),(b) the Hubbard model and (c)-(f) the Hubbard-Holstein model with (c),(d) $\omega_{ph} = 0.25t$, $\lambda = 0.039$ and (e),(f) $\omega_{ph} = t$, $\lambda = 0.12$. The onsets show the density of states in the vicinity of the Fermi level.}
\end{figure}

\begin{figure}
\includegraphics{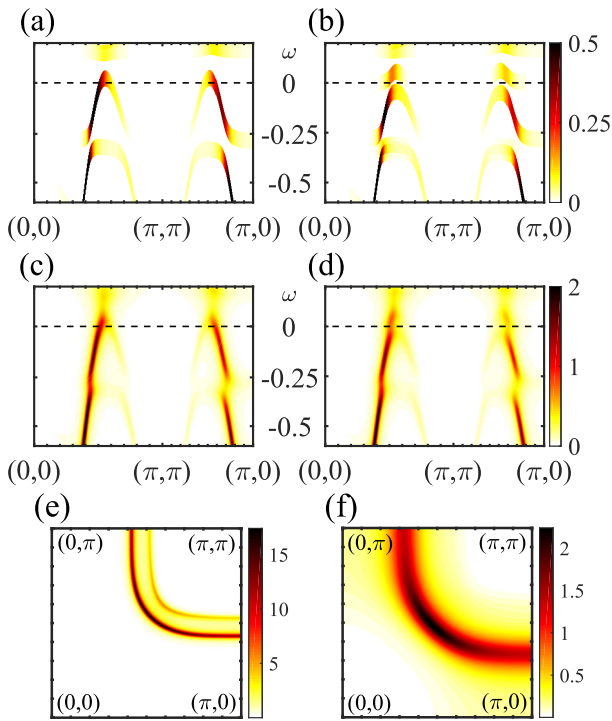}
\caption{\label{fig:Fig8} (a)-(d) $A_{\sigma}\left(\mathbf{k}, \omega\right)$ in the vicinity of the Fermi level and (e),(f) $A_{\sigma}\left(\mathbf{k}, \omega = 0\right)$ within the (a),(c),(e) Hubbard model and (b),(d),(f) Hubbard-Holstein model with $\omega = 0.25t$, $\lambda = 0.039$.  The other model parameters are $p=0.05$, $t'=-0.2t$, $t''=0.15t$, $U=8t$. In (a),(b) no Lorentzian broadening is used, in (c),(d) $\delta = 0.05t$, in (e),(f) $\delta=0.01t$.}
\end{figure}

In general, the effect of electron-phonon interaction on the quasiparticle dispersion curves is their splitting into polaron bands due to a hybridization with Franck-Condon resonances \cite{Sawatzky89} as demonstrated in Ref.~\onlinecite{Makarov15}. For the first set of parameters, comparing the spectral weight distribution and the density of states within the Hubbard model (see Fig.~\ref{fig:Fig7}(a),(b)) and within the Hubbard-Holstein model with a weak electron-phonon interaction $\lambda = 0.039$ (see Fig.~\ref{fig:Fig7}(c),(d)) at modest phonon frequency $\omega_{ph} = 0.25t$   shows this effect at different energy scales of LHB: within the in-gap states slightly above the Fermi level, at energies $-2.5t\lesssim\omega\lesssim-0.5t$, $-4t\lesssim\omega\lesssim-3t$, and also in the vicinity of Fermi level (as seen by comparing Fig.~\ref{fig:Fig8}(a),(b)). In this case a kink-like behavior close to the Fermi level is recognized in Fig.~\ref{fig:Fig8}(b),(d). The Fermi surface is also affected (see \ref{fig:Fig8}(e),(f)) as it is formed by the weak polaronic band when the phonons are present. 

Considering the second set of parameters, $\omega_{ph} = t$, $\lambda = 0.12$, it is seen from comparing panels (a),(b) with (e),(f) of Fig.~\ref{fig:Fig7} that at such a high phonon frequency even moderate electron-phonon coupling $\lambda = 0.12t$ leaves the low-energy electronic structure almost unaffected, while the high-energy structure at $\omega\lesssim-t$ is heavily split into the large number of polaronic bands. It can be interpreted that the high-energy region becomes significantly incoherent. A similar trend can be traced in the data obtained by quantum Monte-Carlo \cite{Nowadnick15}.

\section{\label{sec:Conclusion}Conclusion}
In this paper, CPT have been applied to the Hubbard and Hubbard-Holstein $t-t'-t''-U$ models at low hole doping. Based on full ED of a 3x3 cluster for the Hubbard model and a 2x2 8-phonon cluster for the Hubbard-Holstein model, we have obtained the fine structure of quasiparticle bands of Hubbard fermions and polarons in order to investigate in detail, how different spectral anomalies arise in strongly correlated systems, when short-range correlations from local Coulomb and electron-phonon interactions affect the properties of such quasiparticle bands. Although full ED allows to treat less interactions than the ground state Lanczos- or renormalization group-based \cite{Yang16} cluster solvers, it has an advantage when one is interested in fine features of the electronic structure. The full set of relevant Hubbard quasiparticle bands can be obtained using a norm-conserving approximation without a Lorentzian broadening (which is in fact an additional approximation).

Having analyzed the obtained data on the Hubbard model, we point at the existence of energy scales with qualitatively different properties of quasiparticles. Particularly important is the low-energy single LSW band, which participates in the formation of anomalous spectral features such as hole, electron Fermi pockets, and the feature similar to the low-energy kink. Within our attempt to analyze the electronic structure of the Hubbard-Holstein model with equal account for electron-electron and electron-phonon interaction, we observed that splitting of fermion bands can cause low-energy kink-like features and affect the degree of quasiparticle coherence in the vicinity of the Fermi level at moderate phonon frequency and low electron-phonon interaction. However, the kink-like features from the Coulomb interactions are in general more pronounced. We cannot claim that it proves a pure electronic origin of the low-energy kinks found experimentally, because in cuprates there are many phonon modes interacting with electrons, and we have restricted our consideration by only one mode. At high phonon frequency and moderate electron-phonon interaction we observe a splitting of the high-energy spectrum into a large number of polaron bands, while the low energy part is almost unaffected. 

The results obtained in this paper, due to the construction of the method, arise mainly from strong short-range correlations. In particular, within the Hubbard model we observed the Lifshitz transition at $p\sim0.08$ leading to the Fermi surface with an electron pocket, in agreement with the results from high-field transport measurements, without \textit{ad hoc} introducing of density waves, in the same manner as other anomalous features of the electronic structure.

\begin{acknowledgments}
The reported study was funded by RFBR according to the research project No. 18-32-00256 (all the results except correlation functions calculated using 4x4 clusters). The authors are grateful to the Presidium RAS program No. 12 ``Fundamental problems of high temperature superconductivity'', ``BASIS'' Foundation for Development of Theoretical Physics and Mathematics, and Russian Foundation for Basic Research (grant 16-02-00098), Government of Krasnoyarsk Territory, Krasnoyarsk Region Science and Technology Support Fund to the research project 18-42-243004.
\end{acknowledgments}

\bibliography{paper}

\end{document}